\documentclass[twocolumn,showpacs,aps,prl,superscriptaddress]{revtex4}

\usepackage{graphicx}
\usepackage{dcolumn}
\usepackage{epsfig}
\usepackage{amsmath}

\newcommand{\BaBarYear}    {07}
\newcommand{\BaBarNumber}  {32}

\newcommand{\SLACPubNumber} {12630}

 \newcommand{\BaBarType}      {PUB}  

\input pubboard/babarsym   

\RequirePackage{xspace}

\newcommand{\pvec}{{\bf p}}

\newcommand{\acp}{\ensuremath{\calA_{ch}}}

\newcommand{\calB}{\ensuremath{{\cal B}}}

\newcommand{\DE}{\ensuremath{\Delta E}}

\newcommand{\xf}{\ensuremath{{\cal F}}}
\newcommand{\hel}{\ensuremath{{\cal H}}}
\newcommand{\thetaT}{\ensuremath{\theta_{\rm T}}}
\newcommand{\costhr}{\ensuremath{\cos\thetaT}}

\newcommand\etal{{\it et al.}}
\newcommand{\half}{\ensuremath{\frac{1}{2}}}

\newcommand{\bma}[1]{\boldmath{$#1$}}
\newcommand{\msp}{\ensuremath{\phantom{-}}}

\newcommand{\bfig}{\begin{figure}[htbpc!]}
\newcommand{\efig}{\end{figure}}
\newcommand\bef{\begin{figure}}
\newcommand\edf{\end{figure}}
\newcommand\dbline{\noalign{\vskip 0.10truecm\hrule}\noalign{\vskip 2pt}\noalign{\hrule\vskip 0.10truecm}}
\providecommand{\tbline}{\noalign{\vskip 0.05truecm\hrule\vskip0.05truecm}}

\newcommand\beq{\begin{equation}}
\newcommand\eeq{\end{equation}}
\newcommand\bear{\begin{array}}
\newcommand\enar{\end{array}}
\newcommand\beqa{\begin{eqnarray}}
\newcommand\eeqa{\end{eqnarray}}
\newcommand\ben{\begin{enumerate}}
\newcommand\een{\end{enumerate}}

\newcommand{\UfourS}{\ensuremath{\Upsilon(4S)}}

\newcommand{\etagg}{\ensuremath{\eta_{\gaga}}}
\newcommand{\etappp}{\ensuremath{\eta_{3\pi}}}

\newcommand{\etapepp}{\ensuremath{\etapr_{\eta\pi\pi}}}

\newcommand{\etaprg}{\ensuremath{\etapr_{\rho\gamma}}}

   \newcommand{\rhoz}{\ensuremath{\rho^0}}

\newcommand{\fetapip}{\ensuremath{\eta\pi^+}}
\newcommand{\etapip}{\ensuremath{\Bp\ra\fetapip}}
\newcommand{\Betapip}{\ensuremath{\calB(\etapip)}}
\newcommand{\retapip}{\ensuremath{xx^{+xx}_{-xx}\pm xx}}
\newcommand{\Retapip}{\ensuremath{(\retapip)\times 10^{-6}}}
\newcommand{\Aetapip}{\ensuremath{xx^{+xx}_{-xx}\pm xx}}

   \newcommand{\fetaggpip}{\ensuremath{\eta_{\gaga} \pip}}
   
   \newcommand{\fetappppip}{\ensuremath{\eta_{3\pi} \pip}}

\newcommand{\fetaKp}{\ensuremath{\eta K^+}}
\newcommand{\etaKp}{\ensuremath{\Bp\ra\fetaKp}}
\newcommand{\BetaKp}{\ensuremath{\calB(\etaKp)}}
\newcommand{\retaKp}{\ensuremath{xx^{+xx}_{-xx}\pm xx}}
\newcommand{\RetaKp}{\ensuremath{(\retaKp)\times 10^{-6}}}

\newcommand{\AetaKp}{\ensuremath{xx^{+xx}_{-xx}\pm xx}}

   \newcommand{\fetaggKp}{\ensuremath{\eta_{\gaga} \Kp}}
   
   \newcommand{\fetapppKp}{\ensuremath{\eta_{3\pi} \Kp}}

\newcommand{\fetappip}{\ensuremath{\etapr\pip}}
\newcommand{\etappip}{\ensuremath{\Bp\ra\fetappip}}
\newcommand{\Betappip}{\ensuremath{\calB(\Bp\ra\etapr \pip)}}
\newcommand{\retappip}{\ensuremath{xx^{+xx}_{-xx} \pm xx}}
\newcommand{\Retappip}{\ensuremath{(\retappip)\times 10^{-6}}}

\newcommand{\Aetappip}{\ensuremath{xx\pm xx}}

   \newcommand{\fetapepppip}{\ensuremath{\etapr_{\eta\pi\pi} \pi^+}}
   \newcommand{\etapepppip}{\ensuremath{\Bp\ra\fetapepppip}}
   \newcommand{\fetaprgpip}{\ensuremath{\etapr_{\rho\gamma} \pi^+}}
   \newcommand{\etaprgpip}{\ensuremath{\Bp\ra\fetaprgpip}}

\newcommand{\fetapKp}{\ensuremath{\etapr K^+}}
\newcommand{\etapKp}{\ensuremath{\Bp\ra\fetapKp}}
\newcommand{\BetapKp}{\ensuremath{\calB(\Bp\ra\etapr \Kp)}}
\newcommand{\retapKp}{\ensuremath{xx^{+xx}_{-xx} \pm xx}}
\newcommand{\RetapKp}{\ensuremath{(\retapKp)\times 10^{-6}}}
\newcommand{\AetapKp}{\ensuremath{xx\pm xx}}

   \newcommand{\fetapeppKp}{\ensuremath{\etapr_{\eta\pi\pi} K^+}}
   
   \newcommand{\fetaprgKp}{\ensuremath{\etapr_{\rho\gamma} K^+}}

\newcommand{\fetapKz}{\ensuremath{\etapr K^0}}

\newcommand{\etapKz}{\ensuremath{\Bz\ra\fetapKz}}

\newcommand{\BetapKz}{\ensuremath{\calB(\etapKz)}}
\newcommand{\retapKz}{\ensuremath{xx^{+xx}_{-xx} \pm xx}}
\newcommand{\RetapKz}{\ensuremath{(\retapKz)\times 10^{-6}}}

   \newcommand{\fetapeppKz}{\ensuremath{\etapr_{\eta\pi\pi} K^0}}

   \newcommand{\fetaprgKz}{\ensuremath{\etapr_{\rho\gamma} K^0}}

\newcommand{\fomegapip}{\ensuremath{\omega\pi^+}}
\newcommand{\omegapip}{\ensuremath{\Bp\ra\fomegapip}}
\newcommand{\Bomegapip}{\ensuremath{\calB(\omegapip)}}
\newcommand{\romegapip}{\ensuremath{xx \pm xx\pm xx}}
\newcommand{\Romegapip}{\ensuremath{(\romegapip)\times 10^{-6}}}
\newcommand{\Aomegapip}{\ensuremath{0.xx\pm 0.yy \pm 0.zz}}

\newcommand{\fomegaKp}{\ensuremath{\omega K^+}}
\newcommand{\omegaKp}{\ensuremath{\Bp\ra\fomegaKp}}
\newcommand{\BomegaKp}{\ensuremath{\calB(\omegaKp)}}
\newcommand{\romegaKp}{\ensuremath{xx\pm xx\pm xx}}
\newcommand{\RomegaKp}{\ensuremath{(\romegaKp)\times 10^{-6}}}
\newcommand{\AomegaKp}{\ensuremath{0.xx\pm 0.yy \pm 0.zz}}

\newcommand{\fomegaKz}{\ensuremath{\omega K^0}}
\newcommand{\fomegaKs}{\ensuremath{\omega\KS}}
\newcommand{\omegaKz}{\ensuremath{\Bz\ra\fomegaKz}}
\newcommand{\omegaKs}{\ensuremath{\Bz\ra\fomegaKs}}
\newcommand{\BomegaKz}{\ensuremath{\calB(\omegaKz)}}
\newcommand{\romegaKz}{\ensuremath{5.9^{+1.0}_{-0.9}\pm xx}}
\newcommand{\RomegaKz}{\ensuremath{(\romegaKz)\times 10^{-6}}}

\renewcommand{\retapKp}{\ensuremath{70.0\pm1.5\pm 2.8}}
\renewcommand{\retapKz}{\ensuremath{66.6\pm 2.6\pm 2.8}}

\renewcommand{\romegaKz}{\ensuremath{5.4\pm 0.8\pm 0.3}}

\renewcommand{\romegapip}{\ensuremath{6.7\pm 0.5\pm 0.4}}

\renewcommand{\romegaKp}{\ensuremath{6.3\pm 0.5\pm 0.3}}

\renewcommand{\AetapKp}{\ensuremath{\msp 0.010\pm 0.022\pm 0.006}}
\renewcommand{\Aomegapip}{\ensuremath{-0.02\pm 0.08\pm 0.01}}
\renewcommand{\AomegaKp}{\ensuremath{-0.01\pm 0.07\pm 0.01}}

\renewcommand{\retapip}{\ensuremath{5.0\pm 0.5\pm 0.3}}

\renewcommand{\Aetapip}{\ensuremath{-0.08\pm 0.10\pm 0.01}}

\renewcommand{\retaKp}{\ensuremath{3.7\pm 0.4\pm 0.1}}

\renewcommand{\AetaKp}{\ensuremath{-0.22\pm 0.11\pm 0.01}}

\renewcommand{\retappip}{\ensuremath{3.9\pm 0.7\pm 0.3}}
\renewcommand{\Aetappip}{\ensuremath{0.21\pm 0.17 \pm 0.01}}

\newcommand{\theTitle}{{\boldmath Branching fraction and \CP-violation 
charge asymmetry measurements for $B$-meson decays to 
$\eta K^\pm$, $\eta\pi^\pm$, $\eta^{\prime} K$, $\eta^{\prime}\pi^\pm$, 
$\omega K$, and $\omega\pi^\pm$}

}

\begin{document}

\preprint{\babar-PUB-\BaBarYear/\BaBarNumber}
\preprint{SLAC-PUB-\SLACPubNumber}

\begin{flushleft}
\babar-\BaBarType-\BaBarYear/\BaBarNumber \\
SLAC-PUB-\SLACPubNumber \\
\end{flushleft}


\title{\theTitle}

%
\author{B.~Aubert}
\author{M.~Bona}
\author{D.~Boutigny}
\author{Y.~Karyotakis}
\author{J.~P.~Lees}
\author{V.~Poireau}
\author{X.~Prudent}
\author{V.~Tisserand}
\author{A.~Zghiche}
\affiliation{Laboratoire de Physique des Particules, IN2P3/CNRS et Universit\'e de Savoie, F-74941 Annecy-Le-Vieux, France }
\author{J.~Garra~Tico}
\author{E.~Grauges}
\affiliation{Universitat de Barcelona, Facultat de Fisica, Departament ECM, E-08028 Barcelona, Spain }
% ...
\collaboration{The \babar\ Collaboration}
\noaffiliation

\date{\today}

\begin{abstract}
We present measurements of the branching fractions for \Bz\ meson
decays to \fetapKz\ and \fomegaKz, and of the branching fractions and
\CP-violation charge asymmetries for \Bp\ meson decays to \fetapip,
\fetaKp, \fetappip, \fetapKp, \fomegapip, and \fomegaKp.  The data,
collected with the \babar\ 
detector at the Stanford Linear Accelerator Center, represent 383
million \BB\ pairs produced in \epem\ annihilation. 
The measurements agree with previous results; we find no evidence for direct
\CP\ violation.
\end{abstract}


\maketitle

Charmless $B$ decays are becoming increasingly useful to test the
accuracy of theoretical estimation methods, such as those based on QCD
factorization~\cite{SUthreeQCDFact,acpQCDfact,SCET} or flavor~SU(3)
symmetry~\cite{FUglob,chiangPP,chiangGlob}.  In this paper we present
measurements of branching fractions and, where applicable, charge
asymmetries, for eight charmless $B$ decays (and their charge-conjugates,
implied throughout the paper): \etapip, \etaKp, \etappip,
\etapKp, \etapKz, \omegapip, \omegaKp, and \omegaKz.  The results
presented here
represent improvement in precision over previous measurements of these
quantities by \babar\ \cite{PRLetah05,PRLetapK0td05,PRDRComegaK0td06},
Belle 
\cite{Belle_etah,Belle_etaph,Belle_omegah}\ and CLEO \cite{CLEO}.  
We previously reported a branching fraction limit for $\Bz\ra\eta\Kz$ 
\cite{PRDetaKz}, and \CP\ asymmetries for \etapKz\ and \omegaKz\
\cite{PRLetapK0td07,PRDRComegaK0td06}.

Charmless $B$ decays with kaons are usually expected to be dominated by
$b\ra s$ loop (``penguin") amplitudes, while $b\ra u$ tree amplitudes
typically dominate for the decays with pions.
However, the $\B\ra\eta K$ decays are especially interesting since they
are suppressed relative to the abundant $\B\ra\etapr K$ decays due to
destructive interference between two penguin amplitudes~\cite{Lipkin}.
The Cabibbo-Kobayashi-Maskawa (CKM) suppressed $b\ra u$ tree amplitudes may interfere significantly
with $b\ra s$ penguin amplitudes of similar magnitudes, possibly leading to
large direct \CP\ violation in
\etapip\ and \etappip~\cite{directCP}; numerical estimates are
available in a few cases~\cite{acpQCDfact,SCET,FUglob,acpgrabbag}. We search
for such direct \CP\ violation by measuring the charge asymmetry $\acp
\equiv (\Gamma^--\Gamma^+)/(\Gamma^-+\Gamma^+)$ in the rates
$\Gamma^\pm=\Gamma(B^\pm\ra f^\pm)$ for each charged final state
$f^\pm$.

Finally, phenomenological fits to the branching fractions and charge
asymmetries of charmless $B$ decays can be used to understand the
relative importance of tree and penguin contributions and may provide
sensitivity to the CKM angle
$\gamma$~\cite{FUglob,chiangPP,chiangGlob,soniSuprunGlob}, 
or to the effect of non-Standard-Model heavy particles in the
loops~\cite{beyondSM}.

The results presented here are based on data collected with the \babar\
detector~\cite{BABARNIM} at the PEP-II $e^+e^-$
collider~\cite{pep} located at the Stanford Linear Accelerator Center.
An integrated luminosity of 347 \invfb, corresponding to
$383\times 10^6$ \BB\ pairs, was recorded at the
$\Upsilon (4S)$ resonance (center-of-mass energy $\sqrt{s}=10.58\
\gev$).

Charged particles from the \epem\ interactions are detected, and their
momenta measured, by a combination of five layers of double-sided
silicon microstrip detectors and a 40-layer drift chamber,
both operating in the 1.5~T magnetic field of a superconducting
solenoid. Photons and electrons are identified with a CsI(Tl)
electromagnetic calorimeter (EMC).  Further charged particle
identification (PID) is provided by the average energy loss ($dE/dx$) in
the tracking devices and by an internally reflecting ring imaging
Cherenkov detector (DIRC) covering the central region.

We establish the event selection criteria with the aid of a detailed
Monte Carlo (MC) simulation of the \B\ production and decay sequences,
and of the detector response \cite{geant}.  These criteria are designed
to retain signal events with high efficiency.  When applied to the
data, they result in a sample
much larger than the expected signal, but with well characterized
backgrounds. We extract the signal yields from this sample with a
maximum likelihood (ML) fit.

The \B-daughter candidates are reconstructed through their decays
$\piz\ra\gaga$, $\Kz\ra\KS\ra\pip\pim$, $\omega\ra\pip\pim\piz$,
$\eta\ra\gaga$ (\etagg), 
$\eta\ra\pip\pim\piz$ (\etappp), $\etapr\ra\etagg\pip\pim$ (\etapepp),
and $\etapr\ra\rhoz\gamma$ (\etaprg), where $\rhoz\ra\pip\pim$.  
The invariant mass of these particles' final states are required to lie
within about two standard deviations of the nominal mass
\cite{PDG2006}\ unless the mass 
is an observable in the ML fit, in which case we accept a wider range.
For a \KS\ candidate we require a successful fit of the decay vertex
with the flight
direction constrained to the pion pair momentum direction, yielding a flight
length greater than three times its uncertainty.
Secondary charged pions in \etapr, $\eta$ and $\omega$
candidates are rejected if classified as protons, kaons, or electrons by
their DIRC, $dE/dx$, and EMC PID signatures.  
For the primary charged track in \Bp\ decays we
define the PID variables $S_{\pi}$ and $S_K$ as the number of standard
deviations between the measured DIRC Cherenkov angle and that expected
for pions and kaons, respectively.
We include these observables in the ML fits to distinguish between
primary $\pi$ and $K$. 
For \etapKp\ the backgrounds, including cross feed from the pion
channel, are small.  For this mode we perform a dedicated fit with less
restrictive continuum background rejection (see below), and $S_K<2$ to 
exclude pions (and lighter particles). 

We reconstruct the \B-meson candidate by combining the
four-momenta of a pair of daughter mesons, with a vertex constraint if
the ultimate final state includes at least two charged particles.  Since
the natural widths of the $\eta$, \etapr, and \piz\ are much smaller
than the resolution, we also constrain their masses to nominal values
\cite{PDG2006}\ in the fit of the \B\ candidate.
From the kinematics of \UfourS\ decay we determine the energy-substituted mass
$\mes=\sqrt{\frac{1}{4}s-\pvec_B^2}$
and energy difference $\DE = E_B-\half\sqrt{s}$, where
$(E_B,\pvec_B)$ is the $B$-meson 4-momentum vector, and
all values are expressed in the \UfourS\ frame.
The resolution in \mes\ is $3.0\ \mev$ and in \DE\ is 24--50 MeV, depending
on the decay mode.  We
require $5.25<\mes<5.29\ \gev$ and $|\DE|<0.2$ GeV. 

\begin{figure}[tbp]
\begin{center}
\includegraphics[width=1.\linewidth]{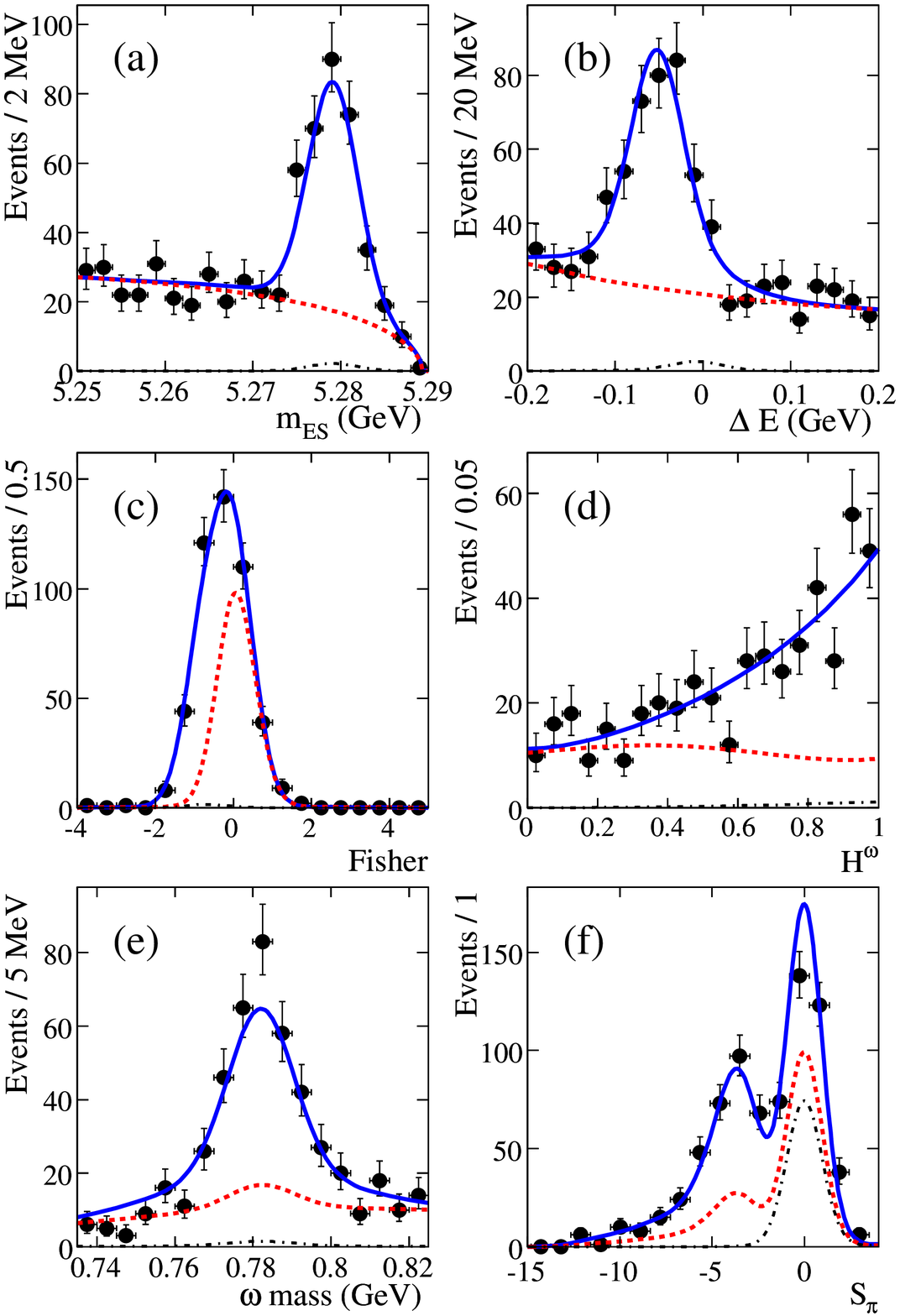}
\caption{Plots of signal-enhanced subsets of the data distribution 
for \omegaKp projected on each
of the fit variables: (a) \mes, (b) \DE, (c) \xf, (d) $\hel^\omega$,
(e) $\omega$ mass, and (f) $S_{\pi}$.  Points with errors represent
the data, solid curves the full fit functions, dashed curves the
sum of the background functions, and dot-dashed curves the signal from
\omegapip.  The variable \DE\ is computed with the pion mass.}
\label{fig:projOmkp}
\end{center}
\end{figure}

Backgrounds arise primarily from random combinations of particles in
continuum $\epem\ra\qqbar$ events ($q=u,d,s,c$).  We reduce these with
requirements on the angle
\thetaT\ between the thrust axis of the $B$ candidate in the \UfourS\
frame and that of the rest of the charged tracks and neutral calorimeter
clusters in the event.  The distribution is sharply
peaked near $|\costhr|=1$ for \qqbar\ jet pairs,
and nearly uniform for $B$-meson decays.  
We require $|\costhr|<0.90$ ($<0.65$ for \fetaprgpip, $<0.80$ for
\fomegapip\ and \fomegaKp),
which optimizes the expected signal yield relative 
to its background-dominated statistical error.
In the ML fit we discriminate further against \qqbar\ background with a
Fisher discriminant \xf\ that combines several variables
which characterize the energy flow in the event \cite{PRD04}.  It
provides about one 
standard deviation of separation between \B\ decay events and \qqbar\
background (see Fig.~\ref{fig:projOmkp}c).

We also impose restrictions on resonance decay angles to exclude the most
asymmetric decays where soft-particle backgrounds accumulate and the
acceptance changes rapidly.  We define the decay angle $\theta_{\rm
dec}^r$ for a meson $r$ that decays to two particles as the angle between
the momenta of a daughter particle and the meson's parent, measured in
the meson's rest frame.  We define $\hel^r\equiv
\cos{\theta_{\rm dec}^r}$ and require $|\hel^{\rhoz}|<0.9$ for
$B\ra\etaprg K$ and $|\hel^{\rhoz}|<0.7$ for \etaprgpip.  For the three-body 
$\omega\ra3\pi$ mode the direction for the decay is the
normal to the decay plane, and we include $\hel^{\omega}$ as an observable in
the ML fit.

The average number of candidates found per 
selected event is in the range 1.05 to 1.13, depending on the final
state.  We choose the candidate with the daughter resonance mass closest 
to the nominal value.  From the simulation we find that
this algorithm selects the correct-combination candidate in about
two thirds of the events containing multiple candidates, and that it
induces negligible bias in the ML fits.

We obtain yields for each channel from an extended maximum likelihood
fit with the input observables \DE, \mes, \xf,
$m_r$ (the invariant mass of the $\eta$,
\etapr, or $\omega$ candidate), and, for charged decays other than
\etapKp, the PID variables $S_{\pi}$ and $S_K$.  
The selected data sample sizes are given in the second column of
Table~\ref{tab:results}.  
Besides the signal events they contain \qqbar\
(dominant) and \BB\ with $b\ra c$ combinatorial background, and a
fraction of background from other charmless \BB\ modes, which we estimate
from the simulation to be less than 2\% of the total fit sample.  The latter 
events have ultimate 
final states different from the signal, but with similar
kinematics so that broad peaks near those of the signal appear in some
observables, requiring a separate component in the probability density
function (PDF).  
The yield of this component is free in the fit for
all cases except 
\omegaKs, where the fit stability requires fixing the yield to the expectation
from MC.  The likelihood function is
\begin{eqnarray}
{\cal L} &=& \exp{\left(\smash[b]{-\sum_{j,k} Y_{jk}}\right)}
\prod_i^{N}\sum_{j,k} Y_{jk} \times \label{eq:likelihood}\\
&&{\cal P}_j (\mes^i) { \cal P}_j(\xf^i) {\cal P}_j (\DE^i_k)\left[
{\cal P}_j (S^i_k){\cal P}_j (m_{r}^i) {\cal P}_j (\hel^i_{r})\right],
\nonumber  
\end{eqnarray}
where $N$ is the number of events in the sample, and for each component
$j$ (signal, combinatorial background, or charmless \BB\ background)
and flavor $k$ (primary \Kp\ or \pip),
$Y_{jk}$ is the yield of events and ${\cal P}_j(x^i_k)$ the
PDF for observable $x_k$ in event $i$.  Some factors in $[\ ]$ are omitted
for some modes.
The flavor-dependent factors ${\cal P}_j (\DE^i_k)$ and ${\cal P}_j
(S^i_k)$ take common functional forms for pion or kaon, e.g.,
$F_j(\DE^i_\pi)$ or
$F_j\left(\DE^i_K=\DE^i_\pi+{\delta\DE}(\pvec^i)\right)$, where \pvec\
is the primary track momentum; $S^i_k$ is treated similarly. 
For the modes $B\ra\etapepp K$ we found no need for the \BB\ background
component.   The factored form of the PDF indicated in Eq.\
\ref{eq:likelihood}\ is a good approximation, particularly for the
\qqbar\ component, since correlations among observables
measured in the data are typically a few percent or less.
Distortions
of the fit results caused 
by our approximations are measured in simulation and
included in the bias corrections and systematic errors discussed below.

We determine the PDFs for the signal and \BB\ background components
from fits to MC samples.  We calibrate the resolutions in \DE\ and
\mes\ with large data control samples of $B$ decays to charmed final states
of similar topology (e.g.\ $B\ra D(K\pi\pi)\pi$).  We develop PDFs for
the combinatorial background with fits to the data from which the signal
region ($5.27\ \gev<\mes<5.29\ \gev$ and $|\DE|<0.1$ GeV) has been excluded.

We use the following functional forms for the PDFs: sum of
two Gaussians for ${\cal P}_{\rm sig}(\mes)$, ${\cal P}_{{\rm sig},\BB}(\DE)$,
and the sharper structures in ${\cal P}_{\BB}(\mes)$ and ${\cal
P}_j(m_r)$; linear or quadratic dependences for 
combinatorial components of ${\cal P}_{\BB,\qqbar}(m_r)$ and for ${\cal
P}_{\qqbar}(\DE)$; and a Gaussian function with separate low- and
high-side width parameters for ${ \cal P}_j(\xf)$.  The \qqbar\  
background in \mes\ is described by the threshold function
$x\sqrt{1-x^2}\exp{\left[-\xi(1-x^2)\right]}$, with
$x\equiv2\mes/\sqrt{s}$ and parameter $\xi$.  These functions are discussed 
in more detail in \cite{PRD04}, and some of them are illustrated in
Fig.~\ref{fig:projOmkp}.  

We allow the parameters most important for the determination of the
background PDFs to vary in the fit, along with the yields
for all components, and for charged modes the signal and \qqbar\
background charge asymmetries.  Specifically, the free background
parameters are most or all of the following, depending on the decay
mode: $\xi$ for \mes, linear and quadratic coefficients for \DE, area
and slope of the combinatorial component for $m_r$, and the mean, width,
and width difference parameters for \xf.  Results for the signal yields are
presented in the third column of Table\ \ref{tab:results}\ for each
sample.

\begin{table*}[btp]
\caption{
Number of events $N$ in the sample, fitted signal yield $Y_S$, and 
measured bias (to be subtracted from $Y_S$) in events (ev.), detection
efficiency $\epsilon$, daughter branching fraction product ($\prod\calB_i$),
and measured branching fraction \calB\ and charge asymmetry \acp\
with statistical error for each
decay chain, and for the combined measurements the 
branching fraction and charge asymmetry with
statistical and systematic error.
The number of produced \BB\ pairs is given in the text.
}
\label{tab:results}
\newcommand{\mn}{\ensuremath{\phantom{-}}}
\newcommand{\eff}{$\epsilon$ (\%)}
\newcommand{\pbf}{$\prod\calB_i$ (\%)}
\newcommand{\signf}{$\cal S$ ($\sigma$)}
\begin{tabular}{lrr@{}lr@{}lcr@{.}lll}
\dbline
Mode	      		& $N$ (ev.)
				&\multicolumn{2}{c}{$Y_S$ (ev.)}
							&\multicolumn{2}{c}{Bias (ev.)}
									&\eff	&\multicolumn{2}{c}{\pbf}
											&\multicolumn{1}{c}{\calB\ $(10^{-6})$}	&\multicolumn{1}{c}{\acp}	\\
\tbline
\bma{\fetapip}		& 	&\multicolumn{2}{c}{}	&\multicolumn{2}{c}{} 	&	&\multicolumn{2}{c}{}&{\boldmath \mn\retapip}	&{\boldmath \Aetapip}	\\ 
~~\fetaggpip		& 44883	&$258$&$^{+30}_{-29}$	&~~~$6$&$\pm$3	& 34.1	&~~~39&4& ~~\mn$4.9\pm0.6$	&~~$-0.05\pm0.12$	\\
~~\fetappppip		& 22333	&$115$&$^{+20}_{-19}$	&$6$&$\pm$3	& 23.8	& 22&6	& ~~\mn$5.5\pm1.0$	&~~$-0.13\pm0.18$	\\
\bma{\fetaKp}		& 	&\multicolumn{2}{c}{}	&\multicolumn{2}{c}{} 	&	&\multicolumn{2}{c}{}&{\boldmath \mn\retaKp}		&{\boldmath \AetaKp}	\\ 
~~\fetaggKp		& 44883	&$197$&$^{+25}_{-24}$	&$6$&$\pm$3	& 32.7	& 39&4	& ~~\mn$3.9\pm0.5$	&~~$-0.25\pm0.13$	\\
~~\fetapppKp		& 22333	&$ 71$&$^{+16}_{-15}$	&$4$&$\pm$2	& 23.2	& 22&6	& ~~\mn$3.3\pm0.8$	&~~$-0.15\pm0.23$	\\
\bma{\fetappip}		& 	&\multicolumn{2}{c}{}	&\multicolumn{2}{c}{} 	&	&\multicolumn{2}{c}{}&{\boldmath \mn\retappip}	&{\boldmath \Aetappip}	\\ 
~~\fetapepppip		& 16879	&$88$&$^{+16}_{-15}$	&$14$&$\pm3$	& 27.2	& 17&5	& ~~\mn$4.0\pm0.9$	&~~$\mn0.14\pm0.20$	\\
~~\fetaprgpip		& 35523	&$97$&$^{+23}_{-22}$	&$23$&$\pm7$& 18.4	& 29&4	& ~~\mn$3.6\pm1.1$	&~~$\mn0.35\pm0.30$	\\
\bma{\fetapKp}		& 	&\multicolumn{2}{c}{}	&\multicolumn{2}{c}{} 	&	&\multicolumn{2}{c}{}&{\boldmath \mn\retapKp}	&{\boldmath \AetapKp}	\\ 
~~\fetapeppKp		&  3170	&$1060$&$\pm35$		&$ 0$&$\pm1$	& 23.2	& 17&5	& ~~\mn$68.2\pm2.3$	&~~$-0.005\pm0.033$	\\
~~\fetaprgKp		& 79501	&$2405$&$\pm69$		&$31$&$\pm16$	& 29.2	& 29&4	& ~~\mn$72.2\pm2.1$	&~~$\mn0.022\pm0.028$	\\
\bma{\fetapKz}		&	&\multicolumn{2}{c}{}	&\multicolumn{2}{c}{} 	&	&\multicolumn{2}{c}{}&{\boldmath \mn\retapKz}	&\multicolumn{1}{c}{(see \cite{PRLetapK0td07})}	\\ 
~~\fetapeppKz		&  1100	&$ 329$&$\pm20$		&$ 3$&$\pm1$	& 23.2	&  6&1	& ~~\mn$60.7\pm3.7$	&\multicolumn{1}{c}{---}	\\ 
~~\fetaprgKz		& 19927	&$ 831$&$\pm38$		&$35$&$\pm17$	& 28.0	& 10&2	& ~~\mn$72.8\pm3.5$	&\multicolumn{1}{c}{---}	\\
\bma{\fomegapip}	& 76735	&$516$&$\pm38$		&$44$&$\pm22$	& 20.5	& 89&1	&{\boldmath \mn\romegapip}	&{\boldmath \Aomegapip}	\\
\bma{\fomegaKp}		& 76735	&$457$&$\pm32$		&$29$&$\pm15$	& 20.0	& 89&1	&{\boldmath \mn\romegaKp}	&{\boldmath \AomegaKp}	\\
\bma{\fomegaKz}		& 15914 &$146$&$\pm18$		&$10$&$\pm5$	& 21.2	& 30&8	&{\boldmath \mn\romegaKz}	&\multicolumn{1}{c}{(see \cite{PRDRComegaK0td06})}	\\
\dbline
\end{tabular}
\end{table*}

We validate the fitting procedure by applying it to
ensembles of simulated \qqbar\ experiments drawn from the PDF into which
we have embedded the expected number of signal and \BB\ background
events randomly extracted from the fully simulated MC samples. 
Biases obtained by this procedure with inputs that reproduce the yields
found in the data are reported in the fourth column of
Table~\ref{tab:results}. 

\begin{figure}
\includegraphics[width=1.\linewidth]{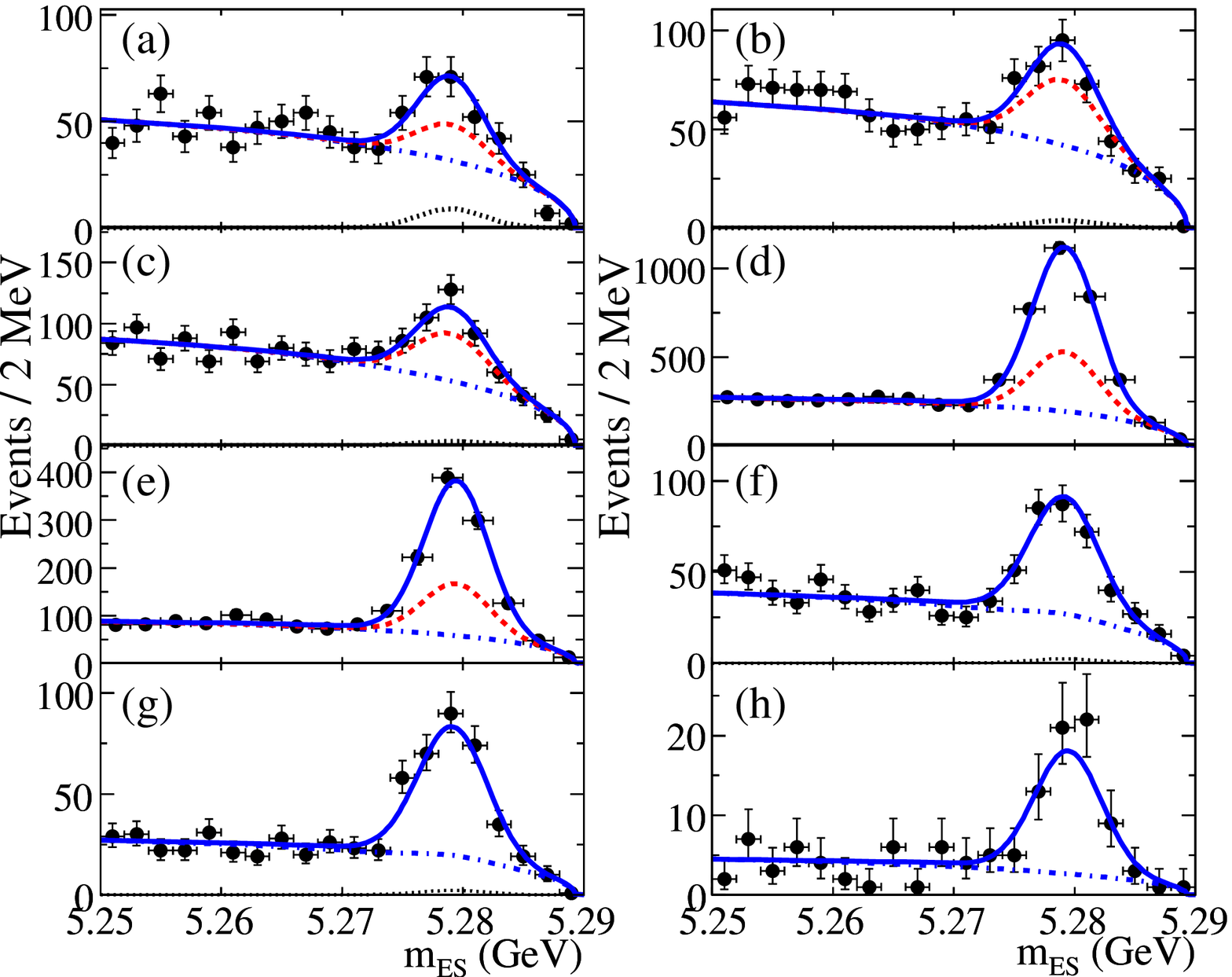}
\caption{\label{fig:proj_mes}Plots of signal-enhanced subsets of the data
distributions projected onto \mes\ for the decays:  (a) 
\etapip, (b) \etaKp, (c) \etappip,
(d) \etapKp, (e) \etapKz, (f) \omegapip, (g) \omegaKp, and (h)
\omegaKz.  The solid line represents the
result of the fit, and the dot-dashed line the background
contribution. 
The dashed line gives the sum of background and the \etappp\ (a, b) or
\etapepp\ (c--e) component of the signal.  The dotted line shows the $K$
or $\pi$ cross-feed component, where applicable.} 
\end{figure}

In Fig.\ \ref{fig:projOmkp}\ we show, as a representative of the
fits, the projections of the PDF and data for the \omegaKp\ fit, and in
Fig.~\ref{fig:proj_mes} projections onto \mes\ for each of the eight
decays, with submodes combined.  The data plotted are subsamples
enriched in signal with a threshold requirement on the ratio of
signal to total likelihood (computed without the plotted variable) that
retains 35\%--80\%\ of the signal, depending on the mode.

We determine the reconstruction efficiencies as the ratio of
reconstructed and accepted events in simulation to the number generated.
We compute the branching fraction for each channel by
subtracting the fit bias from the measured yield, and dividing the
result by the efficiency (including secondary branching fractions) and
the number of produced \BB\ pairs~\cite{PRD04}. 
We assume equal decay rates of the \UfourS\ to \BpBm\ and \BzBzb .
Table \ref{tab:results} gives the numbers pertinent to these computations.
The statistical error on the signal yield or branching fraction is taken
as the change in the central value when the quantity $-2\ln{\cal L}$
increases by one unit from its minimum value. 

We combine results where we have multiple decay channels by adding the
functions $-2\ln{\left\{\left[\calL(\calB)/\calL(\calB_0)\right]\otimes
G(\sigma^\prime)\right\}}$, where $\calB_0$ is the central value
from the fit for each decay channel, and
$\otimes G$ denotes convolution with a Gaussian function to include the
systematic error $\sigma^\prime$ discussed below.  We give the
resulting final branching fractions for each mode in Table
\ref{tab:results}.  

Systematic uncertainties on the branching fractions arise from the PDFs,
\BB\ backgrounds, fit bias, and efficiency.  PDF uncertainties not
already accounted for by free parameters in the fit are estimated from
the consistency of fits to MC and data in control modes.  Varying the
signal-PDF parameters within these errors, we estimate yield
uncertainties of 0.4\%--2.2\%, depending on the mode.
For the \BB\ backgrounds we vary the input
branching fractions within their uncertainties for the modes that
contribute most to the selected sample.  The resulting changes in the
signal yield are taken in quadrature and scaled to the total of all
modes to determine the systematic uncertainty.  For the $\etapepp K$
modes, where no \BB\ component is used, we use 10\% of the expected
\BB\ background in the sample as this is the typical correlation with 
the signal yield. For \fomegaKs\ where the
\BB\ yield is fixed, we take as a systematic uncertainty the average
change in the signal yield when the \BB\ yield is varied between zero
and twice the nominal value.  
The uncertainty of the bias (Table \ref{tab:results}) is a quadrature
sum of its components:  the statistical
uncertainty from the simulated experiments, and half of the corrections
attributable to correlations omitted from the signal and \BB\
background models, and to PID of the primary charged track.  
The primary-track PID correction is significant only for
misidentified kaons from \etapKp\ in the \etappip\ channels.

Uncertainties in our knowledge of the efficiency, found from auxiliary
studies, include $0.5\%\times N_t$ and $1.5\%\times N_\gamma$, where
$N_t$ and $N_\gamma$ are the number of tracks and photons, respectively,
in the $B$ candidate.  The uncertainty in the total number of \BB\ pairs in the
data sample is 1.1\%.  Published data \cite{PDG2006}\ provide the
uncertainties in the $B$-daughter product branching fractions (0.7--3.2\%).
The uncertainties in the efficiency from the event selection are
below 0.5\%.

For the measurements of \acp, biases arise in principle from
charge-dependent effects in the track reconstruction or particle
identification, or from imperfect modeling of the interactions with
material in the detector.  We study these by comparing this effect 
in MC for the
signal, \qqbar\ background in the data, and control samples mentioned
previously.  We apply corrections, and assign systematic errors, to
\acp\ equal to $-0.010\pm0.005$ for modes with a primary kaon and
$0.000\pm0.005$ for those with a primary pion.  We apply an additional
correction with uncertainty for dilution of the \acp\ measurement
associated with the yield bias, which is significant only for
\etapepppip.  This is obtained from the same MC studies that are used to 
estimate the yield bias.

After combining the measurements we obtain
for the branching fractions:
\begin{eqnarray*}
\Betapip &=& \Retapip \\ 
\BetaKp &=& \RetaKp   \\
\Betappip  &=&  \Retappip   \\
\BetapKp  &=&  \RetapKp   \\
\BetapKz  &=&  \RetapKz   \\
\Bomegapip  &=&  \Romegapip   \\
\BomegaKp  &=&  \RomegaKp   \\
\BomegaKz  &=&  \RomegaKz.
\end{eqnarray*}
For the charge asymmetries we find
\begin{eqnarray*}
\acp(\etapip) &=& \Aetapip \\
\acp(\etaKp) &=& \AetaKp  \\ 
\acp(\etappip) &=& \msp\Aetappip \\
\acp(\etapKp) &=& \AetapKp \\
\acp(\omegapip) &=& \Aomegapip \\
\acp(\omegaKp) &=& \AomegaKp.
\end{eqnarray*}
The first error quoted is statistical and the second systematic.
These results are generally consistent with published
measurements
\cite{PRLetah05,PRLetapK0td05,PRDRComegaK0td06,Belle_etah,Belle_etaph,Belle_omegah,CLEO}
and supersede our previous ones
\cite{PRLetah05,PRLetapK0td05,PRDRComegaK0td06}; 
for \BetaKp\ we find a value about twice that of \cite{Belle_etah}.
The theoretical estimates are in agreement with the data (though the data 
have been used in some predictions), but with greater
uncertainty~\cite{SUthreeQCDFact,acpQCDfact,SCET}.
Approaches that fit all available data with a moderate number of model
parameters have proved fruitful \cite{FUglob,chiangPP,chiangGlob}.
We find no clear evidence for direct \CP-violation charge asymmetries in
these decays.  
The world average of the measurements of \acp\ for \etapip\ (\etaKp)
are both negative  and 2.3 (3.0) standard deviations from zero,
while the predictions of \cite{SCET}\ are positive, though with large errors.

We are grateful for the excellent luminosity and machine conditions
provided by our \pep2\ colleagues, 
and for the substantial dedicated effort from
the computing organizations that support \babar.
The collaborating institutions wish to thank 
SLAC for its support and kind hospitality. 
This work is supported by
DOE
and NSF (USA),
NSERC (Canada),
CEA and
CNRS-IN2P3
(France),
BMBF and DFG
(Germany),
INFN (Italy),
FOM (The Netherlands),
NFR (Norway),
MIST (Russia),
MEC (Spain), and
STFC (United Kingdom). 
Individuals have received support from the
Marie Curie EIF (European Union) and
the A.~P.~Sloan Foundation.

\renewcommand{\baselinestretch}{1}

\end{document}